\documentstyle[12pt,preprint,aps,tighten]{revtex}
 
\begin{document}
\input{epsf}
 
\draft
{\tighten

\preprint{
\vbox{
      \hbox{JHU-TIPAC-97016}
      \hbox{UMHEP-446}
      \hbox{hep-ph/9712497} 
      \hbox{December 1997}  }}

\title{Intrinsic charm of light mesons and 
CP violation in heavy quark decay}
\author{Alexey A. Petrov}
\address{
Department of Physics and Astronomy, \\
The Johns Hopkins University, \\ 
Baltimore, MD 21218}
\maketitle
\begin{abstract}
\noindent
We investigate the impact of the intrinsic heavy quark states on the 
predicted values of CP asymmetries in the decays of heavy mesons.
It is shown that the intrinsic charm contribution, although 
dynamically suppressed in QCD, is favored by the weak interaction, 
and therefore it can significantly dilute 
the predicted values of CP-violating asymmetries. This introduces 
additional non-perturbative uncertainty into the estimate of 
direct CP violating effects. We provide a phenomenological estimate
of intrinsic charm content of $\eta$ and $\eta'$ mesons by 
expanding various amplitudes in terms of the heavy-light quark mixing 
angle and discuss theoretical uncertainties in the estimates of 
direct $CP$-violating asymmetries in $B \to \eta^{(\prime)} K^{(*)}$.
\end{abstract}
}
\section{Introduction}
Heavy quark decays offer a wide array of methods of testing 
the Standard Model and searches for the signatures of New Physics.
In particular, they proved to be a powerful tool in the studies of the 
the weak mixing matrix and prepare a fertile ground for the 
exploration of the smallest CKM matrix elements, $V_{ub}$ and $V_{td}$, and,
most importantly, studies of $CP$ violation.
The observation of signatures of direct $CP$ violation involves comparisons of the 
partial decay rates of charged mesons. As a result, the necessary requirement 
is a possibility for two distinct pathways (i.e. amplitudes 
with different weak and strong phases) to reach a given final state. 
It is realized, for instance, in the $B$ decays 
to the final states not containing charmed quarks. In this example,
$CP$ violation can occur either from the interference of the
tree-level and penguin amplitudes, or from the interference of 
the penguin amplitudes with different quark flavors in the loop.
The strong phases are generated by allowing 
internal quarks in the penguin loop to go on their mass shells 
by virtue of the so-called Bander-Silverman-Soni (BSS) 
mechanism \cite{bss}. It is expected that the resulting strong phases for
different amplitudes are different, in one case because tree level graphs 
do not produce perturbative strong phases, and in the other, 
because of the different mass thresholds for different quark species
in the penguin loop.

Let us consider a process governed by the quark $b \to s$
transition to a given final state $f$. In general, there are
two distinct ways to reach $f$: by the tree-level $b \to u \bar u s$
amplitude or by a penguin $b \to s q \bar q$ transition. 
It is clear that the tree-level amplitude is proportional to
the CKM matrix elements $V_{ub}^* V_{us}$ and thus Cabibbo-suppressed 
as it scales as $\lambda^4$ in Wolfenstein parameterization.
On the other hand, the leading penguin effects, although suppressed by
loop factors, scale as $V_{tb}^* V_{ts}$, or $\lambda^2$. This makes the 
tree amplitude comparable in strength with the one-loop penguin diagrams, thus
enhancing the interference term and allowing for sizable 
$CP$-violating asymmetry

\begin{equation}
A^{dir}_{CP} = \frac{\Gamma_{\bar B \to \bar f} - 
\Gamma_{B \to f}}{\Gamma_{\bar B \to \bar f} + 
\Gamma_{B \to f}}. 
\label{asym}
\end{equation}
These asymmetries have been calculated for a number of final states $f$, 
such as $B \to K \bar K, ~K \pi$, etc. 

It is plausible to assume that while BSS mechanism does represent a 
way to produce a non-zero $CP$ violating asymmetry, the soft,
non-perturbative effects might produce larger FSI phases in the 
exclusive transitions, thus introducing a non-perturbative 
uncertainty to the calculation of the asymmetry of Eq. (\ref{asym}). 
For example, the rescattering of physical {\it hadrons} produced in the
reaction provides an additional source for the FSI 
phase \cite{fsi,fknp,neub,ats}.

Here we shall argue that soft FSI contributions do not exhaust 
list of possible non-perturbative uncertainties of $A^{dir}_{CP}$.
The Fock state expansion could include a non-vanishing contribution of the 
heavy quark (e.g. charm) states to light meson's wave function. 
Although higher Fock state contributions are dynamically 
suppressed in QCD, weak transitions of the $b$ quark
to the heavy quark states are Cabibbo-favored. 
Therefore, weak interaction selects intrinsic charm
states of the light mesons making their contributions
competitive with the direct $b \to u$ transitions to the
leading Fock states of the light mesons.
We admit that the intrinsic charm content of light mesons is considerably
difficult to estimate. However, at least for a particular class 
of light vector and pseudoscalar mesons this
contribution can be phenomenologically accounted for by allowing
``mixing'' of the heavy mesons with hidden charm with the light
mesons bearing the same quantum numbers. For instance, there is a 
non-vanishing probability for the mixing of
$J/\psi$ and $\phi$, $\eta_c$ and $\eta'$, etc.
In what follows we consider the possible effects of intrinsic heavy 
quark states of light mesons on $CP$ violating asymmetries.

The paper is organized as follows. In Section II we consider an
upper bound on the value of the heavy-light quark mixing angle.
In Section III we discuss how intrinsic charm content of the light
mesons affects direct CP violating asymmetries concentrating on the
phenomenologically interesting transitions $B \to \eta^{(\prime)} K^{(*)}$.
We summarize our results in Section IV.

\section{Heavy-light quark mixing} 
There are many possible approaches in QCD that account for the intrinsic
heavy quark states in light quark systems. It is therefore 
reasonable to employ a phenomenological description of the 
mixing, extracting the values of the mixing angles from the experiment. 
These values can be later compared to the results obtained using various
models. Let us parameterize a mixing of the heavy and light 
pseudoscalar mesons in terms of the following matrix 

\begin{equation}
\left( \begin{array}{c} 
\eta' \\
\eta \\
\eta_c 
\end{array} \right) 
= \left( \begin{array}{ccc}
a_P-b_P &  c_P-d_P & -\sin \alpha_P \cos (\phi_P-\theta_P) \\
-c_P-d_P & a_P+b_P & -\sin \alpha_P \sin (\phi_P-\theta_P) \\
\sin \alpha_P \cos \phi_P & \sin \alpha_P \sin \phi_P & \cos \alpha_P
\end{array} \right)  ~
\left( \begin{array}{c} 
\eta_0 \\
\eta_8 \\
\eta_{c0} 
\end{array} \right),
\label{mixing}
\end{equation}
where $a_P=\cos^2 (\alpha_P/2) \cos \theta_P$, 
$b_P=\sin^2 (\alpha_P/2) \cos (2 \phi_P-\theta_P)$, 
$c_P=\cos^2 (\alpha_P/2) \sin \theta_P$, and 
$d_P=\sin^2 (\alpha_P/2) \sin (2 \phi_P-\theta_P)$. 
Here $\eta_0$, 
$\eta_8$, and $\eta_{c0}$ represent the flavor SU(3) singlet,
octet and pure $c \bar c$ states respectively, and
$\eta'$, $\eta$, and $\eta_c$ are physical states:
\begin{eqnarray} \label{su3}
|\eta_0 \rangle &=& \frac{1}{\sqrt{3}} |
u \bar u + d \bar d + s \bar s \rangle, \nonumber \\
|\eta_8 \rangle &=& \frac{1}{\sqrt{6}} |
u \bar u + d \bar d - 2 s \bar s \rangle, \\
|\eta_{c0} \rangle &=& |c \bar c \rangle. \nonumber
\end{eqnarray}
The mixing angles $\alpha_P$, $\theta_P$, and $\phi_P$ have a 
simple physical meaning \cite{azimov}:
$\alpha_P$ represents an ``admixture'' of the heavy quarks to
the light ones, $\theta_P$ represents a mixing of the light quarks among 
themselves, and $\phi_P$ gives a light quark $SU(3)$ admixture to 
a heavy $Q \bar Q$ state. It is clear that one recovers a
standard $\eta-\eta'$ mixing matrix as $\alpha_P \to 0$ and
heavy quarks decouple from the light ones.
Sometimes, it is more convenient to work in the quark 
basis\footnote{Note that we are {\it not} using any specific quark model.} 
\begin{eqnarray} \label{qbasis}
|\eta' \rangle &=& \frac{1}{\sqrt{2}} X_{\eta'} |
u \bar u + d \bar d \rangle  + 
Y_{\eta'} | s \bar s \rangle + Z_{\eta'} 
| c \bar c \rangle, \nonumber \\
|\eta \rangle &=& \frac{1}{\sqrt{2}} X_{\eta} |
u \bar u + d \bar d \rangle  + 
Y_{\eta} | s \bar s \rangle + Z_{\eta} 
| c \bar c \rangle, \\
|\eta_c \rangle &=& \frac{1}{\sqrt{2}} X_{\eta_c} |
u \bar u + d \bar d \rangle  + 
Y_{\eta_c} | s \bar s \rangle + Z_{\eta_c} 
| c \bar c \rangle, \nonumber 
\end{eqnarray}
where we have generalized the construction of \cite{gk} to include 
intrinsic charm states. Using Eq. (\ref{mixing}) it is easy to 
show that
\begin{equation}
X_{\eta_i}^2 + Y_{\eta_i}^2 + Z_{\eta_i}^2 = 1
\end{equation}
for each $\eta_i = \{\eta, \eta', \eta_c \}$. This equation might be
violated by the presence of some other pseudoscalars that
might mix with $\eta_i$ as well. This would result in the 
limitation of the described method 
to the prediction of the upper bound on the value of $\eta_c$
contribution to the mixing angle $\alpha_P$.

The construction of Eq.(\ref{mixing}) corresponds to the introduction of the 
additional, charmed singlet current in addition to the ``standard'' 
SU(3) singlet and octet currents
\begin{eqnarray}
A_8^\mu &=& \frac{1}{\sqrt{6}} (\bar u \gamma^\mu \gamma_5 u +
\bar d \gamma^\mu \gamma_5 d - 2 \bar s \gamma^\mu \gamma_5 s),
\nonumber \\
A_0^\mu &=& \frac{1}{\sqrt{3}} (\bar u \gamma^\mu \gamma_5 u +
\bar d \gamma^\mu \gamma_5 d + \bar s \gamma^\mu \gamma_5 s),
\\
A_c^\mu &=& \bar c \gamma^\mu \gamma_5 c.
\nonumber
\end{eqnarray}
These induce the following matrix elements

\begin{eqnarray}
\langle \eta' | \bar u \gamma_\mu \gamma_5 u | 0 \rangle &=&
-i \sqrt{\frac{2}{3}} \left [
(a_P-b_P) F_0 + \frac{1}{\sqrt{2}} (c_P-d_P) F_8 \right ] p_\mu,
\nonumber \\
\langle \eta' | \bar s \gamma_\mu \gamma_5 s | 0 \rangle &=&
-i \sqrt{\frac{2}{3}} \left [
(a_P-b_P) F_0 - \sqrt{2} (c_P-d_P) F_8 \right ] p_\mu,
\\
\langle \eta' | \bar c \gamma_\mu \gamma_5 c | 0 \rangle &=&
i \sqrt{2} \sin \alpha_P \cos (\phi_P - \theta_P) F_{\eta_{c0}}
p_\mu.
\nonumber
\end{eqnarray}
Similar matrix elements exist for the $\eta$.
Here $F_{0,8}$ are the singlet and octet 
decay constants and $F_{\eta_{c0}}$ is $\eta_{c0}$ decay constant.
In the limit of $SU(3)$ symmetry $F_8 = F_\pi$. The $SU(3)$-violating
corrections have been calculated in \cite{dhl} in the framework of
Chiral Perturbation Theory and found to modify this relation by 
approximately $25 \%$, i.e. $F_8/F_\pi \equiv 1/x_8 = 1.25$. We shall 
use this value in the following analysis. The value of $x_0$, on the other 
hand, is not fixed by the $SU(3)$ symmetry arguments, so
we shall keep it as a free parameter fixing it later by fitting to the 
experimental data (In the limit of the ``nonet'' symmetry $F_0/F_\pi \equiv 
1/x_0 = 1$).

The parameters of the mixing matrix can be obtained phenomenologically, 
as they contribute to the decays of charmonia to the light 
$\eta, \eta'$ and $\pi^0$ mesons, and to radiative decays of the 
light mesons. Here our essential assumption is that the mixing angles
$\alpha_P$ and $\phi_P$ are sufficiently small.
This assumption, however, is rather loose, and is valid even for a
relatively large charm content of $\eta'$ \cite{zhitn1,zhitn2}, but it allows us to
carry out a perturbative expansion in these angles.
In the following analysis we will only keep terms linear in
$\alpha_P$ and $\phi_P$.

The bulk of information about the relevant mixing angles comes from the 
radiative decays of $\eta, \eta'$, and $\eta_c$ mesons. Normalizing
the decay widths to the width of $\pi^0 \to \gamma \gamma$ and
using Eq. (\ref{mixing}) we obtain 
\begin{eqnarray} \label{reduced}
\rho_{\eta'} &\equiv& \frac{3}{8} \left ( \frac{m_\pi}{m_{\eta'}}
\right )^3 \frac{\Gamma(\eta' \to \gamma \gamma)}{\Gamma(\pi^0 \to
\gamma \gamma)} \nonumber \\
&=& \left[ \frac{F_\pi}{F_0} \left[ a_P - b_P \right ] +
\frac{F_\pi}{F_8} \frac{c_P - d_P}{\sqrt{8}} -
\sqrt{\frac{3}{8}} \frac{C F_\pi F_{\eta_{c0}}}{m_{\eta_c}^2}
\sin \alpha_P \cos (\phi_P - \theta_P) \right]^2, \nonumber \\
\rho_\eta &\equiv& 3 \left ( \frac{m_\pi}{m_{\eta}}
\right )^3 \frac{\Gamma(\eta \to \gamma \gamma)}{\Gamma(\pi^0 \to
\gamma \gamma)} \nonumber \\
&=& \left[ \frac{F_\pi}{F_0} \left[ a_P + b_P \right] -
\frac{F_\pi}{F_8} \sqrt{8} \left[c_P + d_P \right] -
\frac{1}{\sqrt{3}} \frac{C F_\pi F_{\eta_{c0}}}{m_{\eta_c}^2}
\sin \alpha_P \sin (\phi_P - \theta_P) \right]^2, \\
\rho_{\eta_c} &\equiv& \frac{3}{8} \left ( \frac{m_\pi}{m_{\eta_c}}
\right )^3 \frac{\Gamma(\eta_c \to \gamma \gamma)}{\Gamma(\pi^0 \to
\gamma \gamma)} \nonumber \\
&=& \left[ \frac{F_\pi}{F_0} \sin \alpha_P \cos \phi_P +
\frac{F_\pi}{F_8} \frac{\sin \alpha_P \sin \phi_P}{\sqrt{8}} +
\sqrt{\frac{3}{8}} \frac{C F_\pi F_{\eta_{c0}}}{m_{\eta_c}^2}
\cos \alpha_P \right]^2. \nonumber
\end{eqnarray}
where $a_P$, $b_P$, and $c_P$ were defined in Eq.(\ref{mixing}) and 
\begin{equation}
C=\sqrt{2} \left ( \frac{8 \pi}{3} \right )^2.
\end{equation}
Expanding in terms of $\alpha_P$ and $\phi_P$ and keeping only 
the linear part we arrive at
\begin{eqnarray} \label{pert}
\rho_{\eta'} &\simeq& \left [ \frac{x_8}{\sqrt{8}} \sin \theta_P +
\cos \theta_P \left( x_0 - \sqrt{\frac{3}{8}}
\frac{C F_\pi F_{\eta_{c0}}}{m_{\eta_c}^2} \alpha_P \right)
\right]^2, \nonumber \\
\rho_\eta &\simeq& \left [ x_8 \cos \theta_P -
\sin \theta_P \left( \sqrt{8} x_0 - \sqrt{\frac{1}{3}}
\frac{C F_\pi F_{\eta_{c0}}}{m_{\eta_c}^2} \alpha_P \right)
\right]^2, \\
\rho_{\eta_c} &\simeq& \left [ 
x_0 \alpha_P + \sqrt{\frac{3}{8}}
\frac{C F_\pi F_{\eta_{c0}}}{m_{\eta_c}^2} 
\right]^2. \nonumber
\end{eqnarray}
As it is seen from Eq.(\ref{pert}), the dependence on $\phi_P$ drops
out at this order, so we set $\phi_P=0$ in what follows.
In time, when the accuracy of experimental measurements improves,
the second order in the ``angle expansion'' should constrain the
value of $\phi_P$ as well.
 
It is clear from Eq.(\ref{pert}) that three equations do not
constrain four parameters: $\theta_P, \alpha_P, x_0$, and 
$F_{\eta_{c0}}$. There are several ways to proceed at
this point. For instance, one can fix $x_0 =1$ by assuming
a nonet symmetry and fit the rest of the parameters from the 
Eq.(\ref{pert}). Here we shall take a different approach.
The ``angle expansion'' can be carried out for other
processes involving $\eta$ and $\eta'$, such as 
$J/\psi \to \eta^{(\prime)} \gamma$ and 
$J/\psi \to \eta_c \gamma$, or $\eta' \to \rho \gamma$.
In the limit of SU(3) invariance, only the singlet state
$\eta_0$ or $\eta_{c0}$ can be coupled to $J/\psi$. 
The amplitudes of the radiative decay of a vector charmonium
state into a final state containing the SU(3) singlet light quark
state $|\eta_0 \rangle$  or $| \eta_{c0} \rangle$ can be
written as 
\begin{eqnarray}
A(J/\psi \to \eta_0 \gamma) &=& A \epsilon_{\mu \nu \alpha \beta}
\epsilon_\psi^\mu \epsilon_\gamma^\nu p_\psi^\alpha k_\gamma^\beta,
\nonumber \\
A(J/\psi \to \eta_{c0} \gamma) &=& B \epsilon_{\mu \nu \alpha \beta}
\epsilon_\psi^\mu \epsilon_\gamma^\nu p_\psi^\alpha k_\gamma^\beta.
\end{eqnarray}
Performing the ``angle expansion'' we find that
\footnote{This result agrees with the result of \cite{ag}
in the limit $A/B \ll \alpha_P$. Please note that it is not possible to
extract $\alpha_P$ from this decay mode {\it alone} without invoking
additional dynamical arguments about the size of $A/B$ [{\it cf}.\cite{ag}].}
\begin{eqnarray} \label{psi}
\frac{\Gamma(J/\psi \to \eta_c \gamma)}
{\Gamma(J/\psi \to \eta' \gamma)} &=&
\frac{1}{{\cos}^2 \theta_P} \left[ \frac{p_{\eta_c}}{p_{\eta'}}
\right ]^3 \left[ \frac{1 + \alpha_P (A/B)}{\alpha_P - (A/B)}
\right]^2, \nonumber \\
\frac{\Gamma(J/\psi \to \eta_c \gamma)}
{\Gamma(J/\psi \to \eta \gamma)} &=&
\frac{1}{{\sin}^2 \theta_P} \left[ \frac{p_{\eta_c}}{p_{\eta}}
\right ]^3 \left[ \frac{1 + \alpha_P (A/B)}{\alpha_P - (A/B)}
\right]^2, \\
\frac{\Gamma(J/\psi \to \eta' \gamma)}
{\Gamma(J/\psi \to \eta \gamma)} &=&
\frac{1}{{\tan}^2 \theta_P} \left[ \frac{p_{\eta'}}{p_{\eta}}
\right ]^3. \nonumber
\end{eqnarray}
As seen from Eq.(\ref{psi}), the {\it ratio} of
radiative decay widths of charmonia into $\eta'$ and $\eta$ is
independent of the mixing angle $\alpha_P$ {\it at this order} and
can be used to extract the value of $\theta_P$. Alternatively,
the ratio 
\begin{equation}
\frac{\Gamma(\eta' \to \rho \gamma)}
{3 \Gamma(\omega \to \pi \gamma)} \left [
\frac{p_\pi}{p_\rho} \right]^3 \simeq X_{\eta'}^2 \sim
\frac{1}{3}\left[ \sqrt{2} \cos \theta_P + \sin \theta_P
\right]^2
\end{equation}
does not depend on the heavy-light mixing angle and can be used 
to extract $\theta_P$. Similar analysis is possible for 
$\rho \to \eta \gamma$ etc. 
Extracting $\theta_P$ from either decay mode and feeding it into
Eq.(\ref{pert}) leads to the constraints on all of the 
mixing parameters. This calculation finds $\theta_P \approx -20^o$,
$x_0 \approx 0.92$ (which corresponds to $F_0/F_\pi \approx 1.05$ in full accord 
with \cite{gk,dhl}). 
The situation is more complicated with
respect to the heavy-light mixing angles. Uncertainties in the 
extractions of the light quark mixing angle $\theta_P$ and
decay rates complicate the extraction of the value of the
mixing angle $\alpha_P$. In fact, all of the experimental results can
be successfully fit assuming a zero value for the mixing angle
$\alpha_P$. In order to overcome this problem, additional constraints 
on the parameters in Eq.(\ref{pert}) must be imposed.
For instance, the value of 
$F_{\eta_{c0}}$ can be extracted from other decay modes which
are less sensitive to the heavy-light mixing, e.g. $B \to \eta_c ~X_s$,
where the $c \bar c$ component is enhanced by the CKM matrix element
$V_{cb}$. Using this method, $F_{\eta_{c0}}$ can be fixed to
$F_{\eta_{c0}} \simeq 0.29$ GeV. This and the third line of
Eq.(\ref{pert}) implies that $| \sin \alpha_P | \leq 0.03 ~(1.7^o)$, $\phi_P =0$. 
It is interesting to compare this bound to what already exists in the
literature. It appears that all the models of intrinsic charm 
based on heavy-light meson mixing \cite{ag,acgk} (see also \cite{fk}) satisfy this 
bound. On the other hand, there exists a class of OPE-based calculations 
\cite{zhitn1,zhitn2}
that clearly violates this bound, predicting $\alpha_P \simeq 7^o$ if
the expansion is terminated at the level $1/m_c^2$. As we shall
see later, these values of $\alpha_P$ significantly dilute the
direct CP asymmetries in B-decays\footnote{The OPE-based calculations 
successfully explain the unexpectedly large branching ratio for $B \to \eta' K$.
They however have certain phenomenological difficulties in the case of
the inclusive $\eta'$ production in B-decays, as well
as describing ratios of branching fractions of $B$ meson decaying to
$PP$ and $PV$ final states \cite{ag,kp,ddo}. The fact that
the mass of the charmed quark is not sufficiently large for the fast convergence 
of the OPE might explain the variance in the results of these 
calculations \cite{future}.}.

A similar construction is available for the vector mesons, where
we use the same notations with the obvious replacement
of the subscript $P$ by $V$

\begin{equation}
\left( \begin{array}{c} 
\phi \\
\omega \\
J/\psi 
\end{array} \right) 
= \left( \begin{array}{ccc}
a_V-b_V &  c_V-d_V & -\sin \alpha_V \cos (\phi_V-\theta_V) \\
-c_V-d_V & a_V+b_V & -\sin \alpha_V \sin (\phi_V-\theta_V) \\
\sin \alpha_V \cos \phi_V & \sin \alpha_V \sin \phi_V & \cos \alpha_V
\end{array} \right)  ~
\left( \begin{array}{c} 
\omega_0 \\
\omega_8 \\
\psi_{0} 
\end{array} \right),
\label{vecmix}
\end{equation}
Numerically, 
$\sin \alpha_V \cos (\phi_V - \theta_V) \simeq 1.2 \times 10^{-4}$ 
\cite{azimov}. 

\section{Amplitudes and CP-violating asymmetries}
In $B$-decays, the tree level amplitude is suppressed by a 
small $V_{ub}$, which makes it comparable 
to the one-loop penguin amplitude. On the other hand, the 
transition $B \to [Q \bar Q]s \to Ms$ with Q being a heavy (charm) quark
and $M$ a light final state meson is not CKM suppressed. 
Therefore the latter mechanism becomes competitive
with the direct $b \to u$ transition to the light quarks constituting
the light mesons. Since $b \to c \bar c s$ amplitude does not 
contain a $CP$ violating weak phase, it may significantly 
reduce the predicted value of $CP$ asymmetry. 
In what follows we consider two cases for $M$ being a pseudoscalar 
and a vector meson.

The generic amplitude for the decays of a $B$ meson to the 
charmless final state $f$ can be written as
\begin{equation}
A_{B \to f} = \xi_u A_T + \xi_c A_M + 
\sum_{i=u,c,t} \xi_i A_P^i =
\xi_u \left[ A_T + A_P^{ut} \right] +
\xi_c \left[ A_P^{ct} + A_M \right]. 
\label{amplitude}
\end{equation}
Here $A_T$ is a tree-level amplitude, $A_M$ is a ``mixing''
amplitude and $A_P$ is a penguin amplitude. As usual, unitarity of 
the CKM matrix was used to write $\xi_t = -\xi_c-\xi_u$, 
$A_P^{ct} \equiv A_P^c - A_P^t$, $A_P^{ut} \equiv A_P^u - A_P^t$,
and $\xi_i = V_{ib}^* V_{is}$.

In order to form a $CP$-violating asymmetry, we also must consider
the corresponding amplitude for the decay of the $\bar B$,

\begin{equation}
\bar A_{\bar B \to \bar f} = 
\xi_u^* \left[ \bar A_T + \bar A_P^{ut} \right] +
\xi_c^* \left[ \bar A_P^{ct} + \bar A_M \right]. 
\end{equation}
Using the fact that $\bar A_T = A_T$ and
$\bar A_M = A_M$, the asymmetry of Eq. (\ref{asym}) can be formed as

\begin{eqnarray}
\Delta_f &=& \Gamma_{\bar B \to \bar f} - \Gamma_{B \to f},
\nonumber \\
\Delta_f &=& \lambda_f 
~Im \xi_u^* \xi_c~ Im \left[ A_T + {A_P^{ut}}^* \right]
\left[ A_P^{ct} + A_M \right],
\label{cpva}
\end{eqnarray}
where $\lambda_f = \sqrt{(1-(x_{f_1}+x_{f_2})^2)
(1-(x_{f_1}-x_{f_2})^2)}/(4 \pi m_B)$ 
is a phase space factor, divided by the sum of the corresponding
decay rates.
As seen from Eq.(\ref{cpva}), direct $CP$ violation in the 
$B \to \eta_i K$ mode arise 
not only from the interference of the penguin diagrams with
the different internal quark flavors, but also, due to 
the complicated quark content of the $\eta_i$, from the
interference of the Cabibbo-suppressed tree-level amplitudes
with the penguin amplitudes. As we shall see later,
intrinsic charm contribution affects CP violating asymmetries calculated
using perturbative BSS phases. It is therefore instructive to
study the dependence of the asymmetry on the parameter
$\hat q = q^2/m_B^2$ which parameterizes the momentum flowing 
through the penguin vertex. The final answer, of course, is 
independent of $\hat q$ and should be obtained by either
integrating the asymmetry with respect to $\hat q$ smeared by
some function defined by the momentum distribution of quarks in the
final state mesons, or by fixing $\hat q$ using quark model arguments.
We shall use the second method for our predictions, presenting the
graphs {\it asymmetry vs. $\hat q$} to show the threshold structure
of CP violating asymmetries. We comment on the effects of soft FSI 
phases in the Conclusion.

In the following discussion we shall first use the effective
Hamiltonian calculated at leading order in QCD, i.e.
with no QCD corrections associated with the penguin part. Next,
the full Next-to-Leading order effective Hamiltonian is employed.

It is well-known that the calculation of the two-body non-leptonic 
decays of heavy mesons cannot be performed without invoking 
a particular model. This model dependence is partially cancelled in
the asymmetry Eq.(\ref{asym}). In our calculation we choose 
factorization approximation \cite{fact1,fact2,fact3,fact4,fact5} and
the Bauer, Stech and Wirbel (BSW) model \cite{bsw} to estimate relevant
form-factors.

\subsection{Leading order calculation}
The ``no QCD Hamiltonian'' reads
\begin{eqnarray}
{\cal H}_{eff}^{LO} = \frac{4 G_F}{\sqrt{2}} \Biggl \{
\xi_Q \sum_{i=1}^{2} C_i O_i^Q &-& \frac{\alpha_s}{8 \pi}
\left[ \sum_{i=u,c,t} \xi_i F_i \right] \left [
-\frac{O_3}{N_c} + O_4 - \frac{O_5}{N_c} + O_6 \right ]
\Biggr \}, \\
O_1^q = \bar s_\alpha \gamma_\mu L Q_\beta
\bar Q_\beta \gamma^\mu L b_\alpha~&,&~~
O_2^q = \bar s \gamma_\mu L Q
\bar Q \gamma^\mu L b~,~ \nonumber \\
O_{3(5)} = \bar s \gamma_\mu L b
\sum_q \bar q \gamma^\mu L (R) q~&,&~~ 
O_{4(6)} = \bar s_\alpha \gamma_\mu L b_\beta
\sum_q \bar q_\beta \gamma^\mu L (R) q_\alpha~.~ \nonumber  
\end{eqnarray}
Here, $Q=\{u,c\}$, $q=\{u,d,s\}$, and $F_i$ are the
Inami-Lim functions for the flavor $i$. A similar construction is
available for the effective $b \to d$ transitions (although
the effects of intrinsic charm states are largely suppressed in these modes
since $b \to u \bar u d$ decays are not CKM suppressed compared
to $b \to c \bar c d$). In what follows we drop the contributions 
from the electroweak penguins and dipole operators for the sake of simplicity.

The decay amplitude $A_{\eta^{(\prime)} K} = 
-i \langle \eta^{(\prime)} K| {\cal H}_{eff} |B \rangle$ can be written as

\begin{eqnarray}
A_{B \to \eta^{(\prime)} K} &=& 
\xi_u \left[ A_T^{\eta^{(\prime)} K} + 
F_{ut} A_P^{\eta^{(\prime)} K} \right] +
\xi_c \left[ F_{ct} A_P^{\eta^{(\prime)} K} + 
A_M^{\eta^{(\prime)} K} \right],
\nonumber \\
A_T^{\eta^{(\prime)} K} &=& -G_F m_B^2 
\left[ a_2 F_{\eta^{(\prime)}}^{uu} 
f_+^K(m_{\eta^{(\prime)}}^2) L_k(\mu_i) +
a_1 F_{K} f_+^{\eta^{(\prime)}} (0) L_\eta (\mu_i) \right],
\\
A_P^{\eta^{(\prime)} K} &=& G_F m_B^2 \frac{\alpha_s}{8 \pi} 
\left( 1-\frac{1}{N_c^2} \right) a_P,
\nonumber \\
A_M^{\eta^{(\prime)} K} &=& G_F m_B^2 \sin \alpha_P 
\cos ( \phi_P - \theta_P) F_{\eta_c} 
f_+^K(m_{\eta^{(\prime)}}^2) L_k (\mu_i),
\nonumber
\end{eqnarray}
where $a_1=C_2 + \chi C_1$, $a_2=C_1+ \chi C_2$, $\mu_i=m_i^2/m_B^2$,
and the following notations are used:

\begin{eqnarray}
a_P &=& F_{K} f_+^{\eta^{(\prime)}} (0) L_\eta (\mu_i) +
F_{\eta^{(\prime)}}^{ss} f_+^K(m_{\eta^{(\prime)}}^2) L_k(\mu_i) 
\nonumber \\
&+& 2 F_{K} f_+^{\eta^{(\prime)}} (0) \frac{m_K^2}{m_s m_b}
M_\eta (x_i^2) +
\bar F_{\eta^{(\prime)}}^{ss} f_+^K(m_{\eta^{(\prime)}}^2) 
\frac{m_{\eta^{(\prime)}}^2}{m_s}{m_b} M_k(\mu_i).
\end{eqnarray}
Our choice of the form factors and kinematic parameters is
explained in the Appendix. In the standard factorization approach
$\chi = 1/N_c$ and all the octet-octet non-factorizable corrections 
are neglected. These corrections can be accounted for phenomenologically,
by treating $\chi$ as a free parameter and fitting it to the available data 
assuming universality of these corrections for different final states
\cite{thesis}. In this approach, $a_2 \simeq 0.25$ \cite{ct,cheng}.

In order to maintain gauge invariance and unitarity,
the calculation must be performed to order
$\alpha_s^2$ in the PQCD expansion \cite{gerard}. The 
CP violating asymmetry reads

\begin{eqnarray}
\Delta_{\eta^{(\prime)}K} &=& 
\lambda_{\eta^{(\prime)} K}
~Im \xi_u^* \xi_c ~ \Biggl[
\left (A_P^{\eta^{(\prime)} K} \right )^2 \left \{
Im F_{ut}^* Re F_{ct} +
Re F_{ut}^* \left[ Im F_{ct} - \frac{n_F}{6}
Re F_{ct} \right] \right \} 
\nonumber \\
&+& A_M^{\eta^{(\prime)} K} A_P^{\eta^{(\prime)} K} 
Im F_{ut}^* +
A_P^{\eta^{(\prime)} K} A_T^{\eta^{(\prime)} K} \left[
Im F_{ct} - \frac{n_F}{6}
Re F_{ct} \right] \Biggr].
\end{eqnarray}

\begin{figure}[t]
\centering
\centerline{
\epsfbox{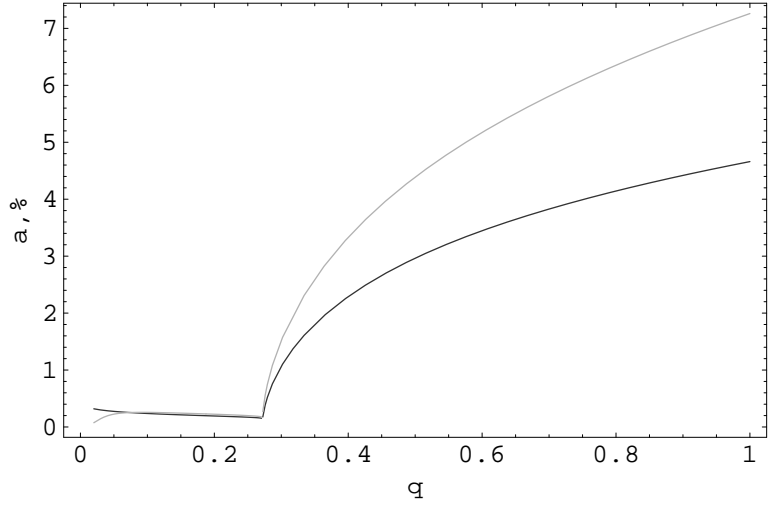}}
\caption{CP asymmetry calculated at the leading order in QCD
as a function of the parameter
$q^2/m_B^2$ for $B \to \eta' K$,
without intrinsic charm (grey) and with intrinsic charm 
$\alpha_P=0.03$ (black).}
\label{scalar}
\end{figure}

\begin{figure}[t]
\centering
\centerline{
\epsfbox{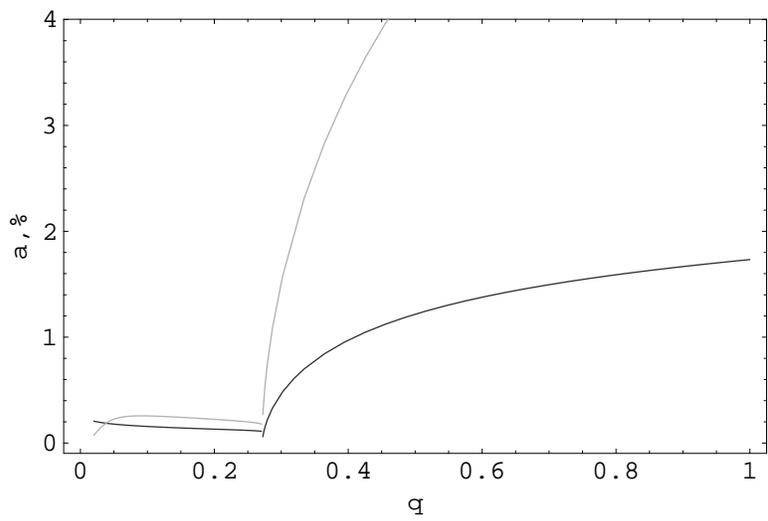}}
\caption{Same as Fig.(1) but for $\alpha_P \simeq 0.12 $ .}
\label{scalarz}
\end{figure}

\noindent
Dividing $\Delta_{\eta' K}$ by the sum of the 
decay rates, the asymmetry Eq. (\ref{asym}) can
be calculated. 
As it is seen from Fig. (\ref{scalar}), 
intrinsic charm reduces CP asymmetry by approximately 
$30-50\%$, if our estimate of $\alpha_P$ is
used thus complicating the extraction 
of CP violating parameters of the CKM matrix from the
decay modes of this type. Please note, that if by some 
reason, the intrinsic charm content of the $\eta'$ is
increased \cite{zhitn1,zhitn2}, this mode becomes practically
useless for the observation of the direct CP violating 
effects. It is clear from the Fig.(\ref{scalarz}) that
CP violating asymmetry is significantly diluted even for
$\alpha_P \simeq 7^o$ which corresponds to the lower bound
of the prediction in OPE-based calculations.
This however, simplifies the time-dependent
analysis of the decay $B^0_d \to \eta^{\prime} K_s$ 
as it suppresses direct CP violating amplitude. The behavior
of CP-violating asymmetries for the case of $\eta K$ final states is
similar to the described above. We therefore refrain from displaying the shape
of $A^{dir}_{CP}(\hat q)$ function, but rather show the numerical value of
$A^{dir}_{CP}$ in Table 1 with $\hat q$ fixed by the quark model arguments. 
The analysis of the $\eta^{(\prime)} V$ final state is completely
similar to the calculation described above. The results are presented 
in Table 1 as well.

\subsection{Next-to-Leading order Calculation}

The NLO QCD effective Hamiltonian reads
\begin{eqnarray}
{\cal H}_{eff}^{NLO} = \frac{4 G_F}{\sqrt{2}} \Biggl \{
\sum_{Q=u,c} \xi_Q & & \Biggl[ \sum_{i=1}^{2} C_i O_i^Q 
+ \sum_{k=3}^{6} C_k O_k  \Biggr ]
\Biggr \}, \\
O_1^q = \bar s_\alpha \gamma_\mu L Q_\beta
\bar Q_\beta \gamma^\mu L b_\alpha~&,&~~
O_2^q = \bar s \gamma_\mu L Q
\bar Q \gamma^\mu L b~,~ \nonumber \\
O_{3(5)} = \bar s \gamma_\mu L b
\sum_q \bar q \gamma^\mu L (R) q~&,&~~ 
O_{4(6)} = \bar s_\alpha \gamma_\mu L b_\beta
\sum_q \bar q_\beta \gamma^\mu L (R) q_\alpha~.~ \nonumber  
\end{eqnarray}
In our calculation we used renormalization scheme independent
effective Wilson's coefficitens defined as
\begin{eqnarray}
C_{2i+1}^{eff~q}(m_b) &=& \bar C_{2i+1}(m_b) + \frac{\alpha_s(m_b)}{8 \pi N_c}
\left ( G (m_q, q^2, m_b) - \frac{10}{9} \right ) \bar C_2, \nonumber \\
C_{2i}^{eff~q}(m_b) &=& \bar C_{2i}(m_b) - \frac{\alpha_s(m_b)}{8 \pi}
\left ( G (m_q, q^2, m_b) - \frac{10}{9} \right ) \bar C_2
\end{eqnarray}
with $i=1,2$ and Wilson's coefficients $\bar C_k$ at $\mu = m_b$ with
$\alpha_s(m_Z)=0.118$ are given by
\begin{eqnarray}
& & \bar C_1 (m_b) = -0.313, ~~ \bar C_2 (m_b) = 1.150, \nonumber \\
& & \bar C_3 (m_b) = 0.017, ~~ \bar C_4 (m_b) = -0.037, \nonumber \\
& & \bar C_5 (m_b) = 0.010, ~~ \bar C_6 (m_b) = -0.046.
\end{eqnarray}
The function $G (m_q, q^2, \mu)$ is given by
\begin{equation}
G (m_q, q^2, \mu) = - 4 \int dx ~ x(1-x) 
\log \frac{m_q^2 - x(1-x) q^2}{\mu^2}
\end{equation}
where $q$ is a momentum of the gluon in the penguin diagram. 
It is clear that the strong phase is generated every time
$q^2 > 4 m_q^2$ for each quark species in the loop.

The factorization calculation is completely similar to the
one performed in the previous section. The only difference arises
from the fact that the effective constants $a_i = C_{2i-1} + \chi C_{2i}$ are not 
vanishing and have to be taken into account along with the corresponding
form-factors. Those are the combinations of the form-factors and
the decay constants defined in the previous section.
The results are presented in Table 1. In our calculation we 
fixed the value of $a_2 \simeq 0.25$ fixed by the experimental data
while dropping the nonfactorizable contributions to the matrix elements of
penguin operators which are difficult to estimate reliably.
The asymmetry is seen not
to change significantly in going to NLO approximation with the 
intrinsic charm contribution still diluting it at the level of 30-50 $\%$
for the estimated value of $\alpha_P \simeq 0.03$. The effect 
becomes stronger for higher values of mixing angles.
As it is seen, even the perturbative result is very uncertain.
The major source of uncertainty is by far dominated by the value of the charmed
quark mass that affects the position of the charmed quark threshold. Because 
of the $SU(3)$ symmetry relations, the values of the hadronic form-factors,
although important for the decay width predictions, are not seen to
significantly affect the predicted values of $CP$ asymmetries. An additional
source of uncertainty is the value of non-factorizable corrections
usually summarized in the effective parameter $\chi$. In the case of the 
asymmetries induced by the penguin-tree interference, it can change the balance 
of these contributions, thus shifting the value of $A^{dir}_{CP}$. It is usually 
assumed that the non-factorizable corrections can be taken into account by 
replacing the factors of $1/N_c$ in $a_i$ throughout the calculation with the 
{\it single} parameter $\chi$. This fact also induces the uncertainty into the 
estimate of both decay width and $CP$-violating asymmetry.\footnote{The contribution 
of the octet matrix element is usually associated with the quark final state 
rescatterings into the colorless hadron states. It is clear that the interactions 
of the rescattered and spectator quarks might introduce different $\chi$'s for 
different amplitudes. For instance, there would be different contributions from
$b \to u \bar u s$ and $b \to d \bar d s$ octet amplitudes to 
$B^- \to \eta^{(\prime)} K^{(*)}$. The effect is similar to the effect of Pauli
interference in nuclear physics.} It is interesting to note that in the limit
$\chi \to 0$ the contribution from the intrinsic charm amplitude
becomes {\it negative} actually reducing the predicted values of branching 
fractions: for instance, $Br(B^- \to \eta' K^-)=2.8 \times 10^{-5}$ for
moderate values of the form-factors and $\alpha_P \simeq 0.3$ drops to
$Br(B^- \to \eta' K^-)=1.6 \times 10^{-5}$ if $\chi = 0$. Of course,
higher values for the branching fraction are still possible considering 
large uncertainties in the values of hadronic form-factors \cite{kp}.
  
Similar calculation can be performed for the 
decays involving vector particles in the final state,
such as $B \to \phi K$. The calculation is 
simplified considerably since the decay is
dominated by a single penguin amplitude. The CP asymmetry for
the leading order case is 
\begin{equation}
\Delta_{\phi K} =
\lambda_{\phi K}
~Im \xi_u^* \xi_c ~ \Biggl[
{A_P^{\phi K}}^2 \left \{
Im F_{ut}^* Re F_{ct} +
Re F_{ut}^* Im F_{ct}  \right \} + 
A_M^{\phi K} A_P^{\phi K} Im F_{ut}^* 
\Biggr].
\label{kphias}
\end{equation}
Here, the following notations are used
\begin{eqnarray}
A_P^{\phi K} &=& - G_F m_\phi F_\phi (p_B + p_K) \cdot \epsilon
~f_+^K(m_{\phi}^2) \frac{\alpha_s}{8 \pi} \nonumber \\
A_M^{\phi K} &=& - G_F m_\phi F_\psi 
(p_B + p_K) \cdot \epsilon ~f_+^K(m_{\phi}^2) a_2
\sin \alpha_V \cos (\phi_V - \theta_V)
\end{eqnarray} 
As usual, one expects partial cancelations among the
first and the second term in Eq.(\ref{kphias}) because 
of the GIM mechanism. Thus, the intrinsic charm 
amplitude is potentially important as it does not
suffer from this cancellation. Fortunately, the small value
of the heavy-light mixing angle for the vector mesons 
makes the intrinsic charm contribution extremely small.
The correction to the asymmetry $A^{dir}_{CP} \simeq 0.3 ~\%$
is less than $1~\%$ for the same choice of CKM parameters
as before. We therefore see no point in 
performing the NLO studies of this class of decay modes.
Clearly, final states containing vector particles are much 
less affected by higher Fock state ``pollution''. The resulting
CP violating asymmetries, however, are significantly smaller.

\section{Conclusions and outlook}
We have investigated the impact of the intrinsic heavy quark
states on the CP violating asymmetries in B-decays. 
It arises due to the fact that the intrinsic charm quark states, 
although suppressed by QCD dynamics of the process, are
less affected by weak Cabibbo suppression. The most dramatic 
effect is seen to occur in the case of the $\eta' V$ final states.
Unfortunately, the impact of the intrinsic charm states on the CP 
asymmetry is very difficult to test experimentally 
since the direct CP asymmetry explicitly
depends on the strong phase of $S$-matrix element relating two
states produced by the weak interaction. 
This quantity is notoriously difficult to 
estimate theoretically since it comprises not only perturbative 
BSS phases but also soft, nonperturbative
phases generated by the rescattering of physical hadrons \cite{fsi}.
Soft FSI contributions, although important, are not seen to 
change the shape of the graphs {\it asymmetry vs. parameter q} but 
rather move them up or down, whereas the described mechanism 
certainly affects the shape. Of course, in the final result the 
function described by $A^{dir}_{CP}(\hat q)$ has to be integrated over 
$\hat q$ with
a suitable choice of "smearing function" representing momentum
distribution of quarks inside the final hadrons, or $\hat q$ has to
be fixed by the quark model arguments. In either case, the intrinsic
charm contribution is seen to reduce the value for CP violating
asymmetry sizably. This effect is not universal, but rather
specific to the final states containing $\eta^{(\prime)}$ 
mesons, so it cannot be attributed to the long-distance part
of the penguin diagram.

The mechanism described in this paper, along with the uncertainty 
in the determination of FSI phases complicates the use of 
the direct CP violating asymmetries as a so-called
``consistency check'' in the determination of the angles 
of the CKM triangle. For instance, the numerical values of 
the angle $\gamma$ determined from various decays modes 
have to be consistent with each other unless there exists a 
New Physics contribution that violates this requirement. 
Thus, the possible inconsistency should manifest a Non-Standard 
Model mechanism affecting the decay processes. As one can
see from the discussion above, there exist possible
sources of violation of this consistency check {\it within}
the Standard Model.

Finally, we would like to note that this mechanism does not markedly 
affect the decay modes where electroweak (EW)
penguins are manifest (e.g. $B \to \eta' \pi$) since there the
tree-level and mixing amplitudes contribute at the same order in 
Wolfenstein parameter $\lambda$ with mixing amplitudes additionally
suppressed by the small values of heavy-light mixing angles.

\acknowledgements
I would like to thank Gene Golowich, Barry Holstein, Nathan Isgur, Alex Kagan,
Amarjit Soni, Daniel Wyler and especially John Donoghue and Adam Falk
for useful discussions and insightful comments.
This work was supported by the National Science Foundation under 
Grants Nos. PHY-9404057 and PHY-9457916, and by the Department of
Energy Grant No. DE-FG02-94ER40869.

\appendix

\section{Hadronic matrix elements and kinematical factors}

We use the following definitions for the
pseudoscalar hadronic form-factors
\begin{eqnarray}
\langle M | \bar q \gamma_\mu b | B \rangle &=&
f_+^M (q^2) (p_B+p_M)_\mu + f_-^M (q^2) (p_B-p_M)_\mu,
\nonumber \\
\langle M | \bar q_1 \gamma_\mu \gamma_5 q_2 | 0 \rangle &=&
-i \sqrt{2} F_M^{q_1 q_2} {p_M}_\mu.
\end{eqnarray}
We take $F_K=0.12~GeV$. Usual relations among decay constants (and transition
form-factors) of different mesons in the pseudoscalar octet imposed by
$SU(3)$-symmetry are modified in the presence of $\eta-\eta'$ mixing and
intrinsic charm components of $\eta $ and $\eta'$, e.g.
\begin{eqnarray}
F_{\eta'}^{uu} &=& \frac{1}{\sqrt{3}} \left \{
(a_P-b_P) F_0 + \frac{1}{\sqrt{2}} (c_P-d_P) F_8 \right \}, \nonumber \\
F_{\eta'}^{ss} &=& \frac{1}{\sqrt{3}} \left \{
(a_P-b_P) F_0 - \sqrt{2} (c_P-d_P) F_8 \right \}, \nonumber \\
f_+^{\eta'} &=& \frac{1}{\sqrt{3}} f_+^0 \left \{
(a_P-b_P) + \frac{1}{\sqrt{2}}(c_P-d_P) \right \},
\end{eqnarray}
where $F_0 \simeq 1.05 F_\pi, F_8 \simeq 1.25 F_\pi$ as explained in the text, 
and $F_\pi = 0.093~GeV$. Also, $f_+^0 = 0.33$. Similar expressions exist for the 
$\eta$-meson as well.
Vector and axial vector form-factors are defined as
\begin{eqnarray}
\langle K^* | \bar s \gamma_\mu (1+\gamma_5) b | B \rangle &=&
i g (q^2) \epsilon_{\mu \nu \alpha \beta} \epsilon^{* \nu}
(p_B + p_K)^\alpha (p_B - p_K)^\beta +
(m_B^2 - m_K^2) f_1 (q^2) \epsilon^*_\mu 
\nonumber \\
&+& (\epsilon^* \cdot q) \left [ f_2 (q^2) (p_B+p_K)_\mu + 
f_3 (q^2) (p_B-p_K)_\mu \right ], \nonumber \\
\langle M^* | \bar q_1 \gamma_\mu q_2 | 0 \rangle &=&
-i \sqrt{2} F_{M^*} m_{M^*} \epsilon^*_\mu.
\end{eqnarray}
The relationship of the form-factors used in this paper to those of
\cite{bsw} is
\begin{eqnarray}
g (q^2) &=& \frac{i V (q^2)}{m_B + m_{K^*}}, ~
f_1 (q^2) = \frac{i A_1 (q^2)}{m_B - m_{K^*}}, ~
f_2 (q^2) = \frac{i A_2 (q^2)}{m_B + m_{K^*}}, \nonumber \\
f_3 (q^2) &=&  \frac{i}{q^2} \left[
2 m_{K^*} A_0 - (m_B + m_{K^*}) A_1 + (m_B - m_{K^*}) A_2
\right ]
\end{eqnarray}
where we take $A_1=A_2=A_3=0.33~GeV$ for the sake of simplicity.
$L(\mu_i)$ and $M$ are the kinematical parameters:

\begin{eqnarray}
L_k(\mu_i) &=& 1- \mu_K + \frac{f_-^K (q^2)}{f_+^K (q^2)} 
\mu_\eta, \nonumber \\
L_\eta(\mu_i) &=& 1- \mu_\eta + \frac{f_-^\eta (q^2)}{f_+^\eta (q^2)} 
\mu_K, \nonumber \\
M_k(\mu_i) &=& \frac{1}{2} \Biggl [(3-y+(1-3y)\mu_K -(1-y)\mu_\eta)
\nonumber \\
&+& \frac{f_-^K (q^2)}{f_+^K (q^2)} 
(1-y+(1-y)\mu_K +(1+y)\mu_\eta)
\Biggr],
\end{eqnarray}
and $y \simeq 1/2 \div 1$ is related to the momentum distribution of quarks
inside of the mesons.

\begin{table}
\begin{center}
\caption{CP-Violating Asymmetries $A^{dir}_{CP}$ for 
$\hat q = q^2/m_b^2 \simeq 0.5$, $\rho=0.05$, $\eta=0.36$,
and $m_c = 1.6 \div 1.3~~GeV$ \label{one}}
\vspace*{0.5cm} 
\begin{tabular}{|l|c|c|c|}
\hline
\phantom{qwerty} Mode & $A^{dir}_{CP} (\alpha_P=0.0),~\%$ \phantom{q} & 
$A^{dir}_{CP}(\alpha_P=0.03),~\%$ \phantom{q} &
$A^{dir}_{CP}(\alpha_P=0.12),~\%$ \phantom{q} \\ 
\hline
\phantom{qwer} $B^- \to \eta' K^-$, LO \phantom{qwe}
& $1.9 \div 3.5$ & $1.6 \div 2.7$ & $1.0 \div 1.5$ \\
\phantom{qwer} $B^- \to \eta' K^-$, NLO  \phantom{qw}
& $2.5 \div 4.4$ & $1.7 \div 2.9$ & $0.7 \div 1.2$ \\
\phantom{qwer} $B^- \to \eta K^-$, LO  \phantom{qwe}
& $3.1 \div 7.3$ & $2.7 \div 6.1$ & $1.9 \div 3.9$ \\
\phantom{qwer} $B^- \to \eta K^-$, NLO \phantom{qw}
&  $3.0 \div 7.1$  & $2.2 \div 5.4$ & $0.8 \div 2.6$ \\
\phantom{qwer} $B^- \to \eta' K^{*-}$, LO \phantom{qwe}
&  $2.9 \div 6.7$ & $1.9 \div 4.1$ & $0.9 \div 1.6$ \\
\phantom{qwer} $B^- \to \eta' K^{*-}$, NLO \phantom{qw}
&  $7.6 \div 16.5$ & $3.3 \div 7.5$ & $0.3 \div 1.3$ \\
\phantom{qwer} $B^- \to \eta K^{*-}$, LO \phantom{qwe}
& $6.7 \div 19.8$ & $4.5 \div 12.5$ & $1.9 \div 4.7$ \\
\phantom{qwer} $B^- \to \eta K^{*-}$, NLO \phantom{qw}
& $15.4 \div 38.3$ & $6.3 \div 18.3$ & $1.4 \div 1.9$ \\
\hline
\end{tabular}
\end{center}
\end{table}

\end{document}